\PassOptionsToPackage{unicode=true}{hyperref} 
\PassOptionsToPackage{hyphens}{url}
\documentclass[12 pt,]{article}
\usepackage{lmodern}
\usepackage{amssymb,amsmath}
\usepackage{ifxetex,ifluatex}
\usepackage{fixltx2e} 
\ifnum 0\ifxetex 1\fi\ifluatex 1\fi=0 
  \usepackage[T1]{fontenc}
  \usepackage[utf8]{inputenc}
  \usepackage{textcomp} 
\else 
  \usepackage{unicode-math}
  \defaultfontfeatures{Ligatures=TeX,Scale=MatchLowercase}
\fi
\IfFileExists{upquote.sty}{\usepackage{upquote}}{}
\IfFileExists{microtype.sty}{%
\usepackage[]{microtype}
\UseMicrotypeSet[protrusion]{basicmath} 
}{}
\IfFileExists{parskip.sty}{%
\usepackage{parskip}
}{
\setlength{\parindent}{0pt}
\setlength{\parskip}{6pt plus 2pt minus 1pt}
}
\usepackage{hyperref}
\hypersetup{
            pdftitle={A pragmatic adaptive enrichment design for selecting the right target population for cancer immunotherapies},
            pdfauthor={Anh Nguyen Duc, Dominik Heinzmann, Claude Berge and Marcel Wolbers},
            pdfborder={0 0 0},
            breaklinks=true}
\usepackage[margin=1in]{geometry}
\usepackage{listings}

\usepackage{graphicx,grffile}
\makeatletter
\def\maxwidth{\ifdim\Gin@nat@width>\linewidth\linewidth\else\Gin@nat@width\fi}
\def\maxheight{\ifdim\Gin@nat@height>\textheight\textheight\else\Gin@nat@height\fi}
\makeatother
\setkeys{Gin}{width=\maxwidth,height=\maxheight,keepaspectratio}
\providecommand{\tightlist}{%
  \setlength{\itemsep}{0pt}\setlength{\parskip}{0pt}}
\ifx\paragraph\undefined\else
\let\oldparagraph\paragraph
\renewcommand{\paragraph}[1]{\oldparagraph{#1}\mbox{}}
\fi
\ifx\subparagraph\undefined\else
\let\oldsubparagraph\subparagraph
\renewcommand{\subparagraph}[1]{\oldsubparagraph{#1}\mbox{}}
\fi

\makeatletter
\def\fps@figure{htbp}
\makeatother

\usepackage{docmute,amsmath} \usepackage{booktabs} \usepackage{longtable} \usepackage{array} \usepackage{multirow} \usepackage{wrapfig} \usepackage{float} \usepackage{colortbl} \usepackage{pdflscape} \usepackage{tabu} \usepackage{threeparttable} \usepackage{threeparttablex} \usepackage[normalem]{ulem} \usepackage{makecell} \usepackage{xcolor} \usepackage[super, sort&compress]{natbib} \setcitestyle{comma,numbers,super,open={},close={}} \makeatletter \renewcommand\@biblabel[1]{#1.} \makeatother
\usepackage{booktabs}
\usepackage{longtable}
\usepackage{array}
\usepackage{multirow}
\usepackage{wrapfig}
\usepackage{float}
\usepackage{colortbl}
\usepackage{pdflscape}
\usepackage{tabu}
\usepackage{threeparttable}
\usepackage{threeparttablex}
\usepackage[normalem]{ulem}
\usepackage{makecell}
\usepackage{xcolor}

\title{A pragmatic adaptive enrichment design for selecting the right target
population for cancer immunotherapies}
\author{Anh Nguyen Duc, Dominik Heinzmann, Claude Berge and Marcel Wolbers}
\date{09 June, 2020}

\begin{document}
\maketitle
\begin{abstract}
One of the challenges in the design of confirmatory trials is to deal
with uncertainties regarding the optimal target population for a novel
drug. Adaptive enrichment designs (AED) which allow for a data-driven
selection of one or more pre-specified biomarker subpopulations at an
interim analysis have been proposed in this setting but practical case
studies of AEDs are still relatively rare. We present the design of an
AED with a binary endpoint in the highly dynamic setting of cancer
immunotherapy. The trial was initiated as a conventional trial in early
triple-negative breast cancer but amended to an AED based on emerging
data external to the trial suggesting that PD-L1 status could be a
predictive biomarker. Operating characteristics are discussed including
the concept of a minimal detectable difference, that is, the smallest
observed treatment effect that would lead to a statistically significant
result in at least one of the target populations at the interim or the
final analysis, respectively, in the setting of AED.
\end{abstract}

\hypertarget{introduction}{%
\section{Introduction}\label{introduction}}

Cancer immunotherapy (CIT) has revolutionized the treatment of cancer
patients. CIT is able to stimulate and promote the immune system and to
engage it in the fight against cancer. The immune system normally
recognizes and eliminates most early tumor cells, but immunological
checkpoints (e.g.~PD-L1) constitute a significant obstacle to effective
antitumor immune responses \citep{Chen_and_Mellman_2013}. An important
class of CITs are PD1 and PD-L1 inhibitors
\citep{Chen_and_Mellman_2013}. PD-L1 protein expression on tumor or
immune cells has emerged as a potential predictive biomarker for
sensitivity to such CITs. However, uncertainty remains on the value of
PD-L1 as a predictive biomarker which may vary by cancer type or stage
\citep{Davis_and_Patel_2019}, as immune-based interactions are dynamic
and complex in nature.

A major topic in research and development of CITs is thus the
identification and confirmation of subgroups of patients where a
treatment is (most) effective \citep{George_et_al_2019} and dealing with
uncertainties regarding the optimal target population is an important
consideration in the design of pivotal CIT trials. If there is
confidence at the time of the design of the pivotal trial that the
biomarker has the ability to identify patients who will benefit from the
treatment, then the trial can be limited to the biomarker-positive
patients (that is, the trial is enriched at the beginning). If the
biomarker is appropriately developed but confidence in its ability to
fully identify the correct biomarker population is lacking, then a
separate Phase II study investigating the biomarker population could be
conducted to inform the Phase III trial design. Alternatively, a single
confirmatory adaptive enrichment design (AED) could be conducted which
allows a data-driven selection of one or more pre-specified biomarker
subpopulations at an interim analysis, and the confirmatory proof of
efficacy in the selected subset at the end of the trial
\citep{Wang_2007,Kaspar2016}.

Regulatory guidance documents for confirmatory adaptive designs are
available and stress the importance of prospective planning of
adaptations and strong type I error control \citep{FDA_AD,EMA_AD}.
Moreover, an ICH guideline \citep{ICH_E20} on adaptive trials is in
preparation. However, although the methodological foundation for
adaptive designs was established more than 30 years ago, their impact in
drug development has not been as high as anticipated
\citep{Peter_Bauer_2015}. In particular, practical case studies of AEDs
are still rare. For example, according to a recent review of 59
medicines for which an adaptive clinical trial had been submitted to the
EMA Scientific Advice, only 5/59 (8\%) concerned AED and for only one of
them, it could be established that the corresponding trial was actually
initiated \citep{Collignon_2018}.

In this article, we present a case study of a confirmatory AED in CIT.
This trial compares a CIT (atezolizumab) plus chemotherapy versus
chemotherapy alone in early triple-negative breast cancer (TNBC) with
pathological complete response as the primary (binary) endpoint. The
trial was originally planned as a conventional randomized trial in
all-comers but then converted to an AED based on emerging data external
to the trial suggesting that PD-L1 could be a predictive biomarker for
the atezolizumab treatment effect in TNBC. The remainder of this article
is structured as follows. Section 2 outlines the general methodological
framework for an AED with a binary clinical endpoint. This section also
contains a discussion of the minimal detectable difference (MDD), an
important concept for the design of a trial. In Section 3, the
methodology is applied to our case study. The article concludes with a
discussion.

\hypertarget{adaptive-enrichment-designs-with-binary-endpoints}{%
\section{Adaptive enrichment designs with binary
endpoints}\label{adaptive-enrichment-designs-with-binary-endpoints}}

\hypertarget{general-description}{%
\subsection{General description}\label{general-description}}

At the design stage of a trial, it is often uncertain whether all
patients or only a targeted subgroup will benefit from the experimental
treatment. Adaptive enrichment designs (AED) which allow for a
data-driven population selection at interim analyses have been proposed
in this setting \citep{Wang_2007,jenkins2010}. The general methodology
to control for multiple testing in such designs via \(p\)-value
combination tests and the closed testing principle was described in
Brannath et al \citep{Brannath_2009}.

In this section, we summarize the theory of AEDs with a focus on methods
relevant to our case study which is a two-stage AED with two
sub-populations defined by a dichotomized biomarker and a binary
endpoint (responder vs non-responder). We refer to Section 11.2 of
Wassmer and Brannath \citep{Wassmer2016a} for a detailed discussion of
the general case with multiple stages and sub-populations and general
endpoints and to \citep{Thomas_2016} for a systematic review of methods
for identification and confirmation of targeted subgroups in clinical
trials. In the last part of this section, we discuss the calculation of
the minimum detectable difference (MDD), which translates the local
significance levels to the clinically more interpretable treatment
effect scale. To our knowledge the MDD has not been discussed in the AED
setting previously.

Let \(F\) denote the full population, \(S\) the sub-population of
subjects tested positive for a binary biomarker of interest, and \(C\)
the subgroup of biomarker-negative subjects. The true response
probability in the experimental arm in \(S\) and \(F\) is denoted by
\(\pi_{1}^q\) with \(q\in\{F,S\}\) and the corresponding probability in
the control arm is \(\pi_{2}^q\). The two one-sided null-hypotheses of
interest are \(H_{0,q}:\pi_{1}^q-\pi_{2}^q\le 0\) versus the
alternatives \(H_{A,q}:\pi_{1}^q-\pi_{2}^q >0\) for \(q\in\{F,S\}\).

A flow chart of a pivotal AED in this setting is shown in Figure 1. In
stage 1 of the trial, \(n_1\) patients from the full population are
randomized to the experimental treatment or control and \(F\) and \(S\)
are co-primary populations. After data from stage 1 is available, a
pre-specified interim analysis is conducted and one of the following
decisions is taken based on decision criteria as discussed in Section
2.3:

\begin{enumerate}
\def\labelenumi{\arabic{enumi}.}
\item
  Stop early for efficacy in \(F\) or \(S\) or both.
\item
  Stop early for futility in both \(F\) and \(S\).
\item
  Continue to stage 2 with \(S\) as the target population, that is,
  randomize an additional \(n_2\) patients from \(S\) in stage 2, and
  only test \(H_{0,S}\) at the final analysis.
\item
  Continue to stage 2 with \(F\) as the target population, that is,
  randomize an additional \(n_2\) patients from \(F\) in stage 2, and
  only test \(H_{0,F}\) at the final analysis.
\item
  Continue to stage 2 with \(F\) and \(S\) as co-primary populations,
  that is, randomize an additional \(n_2\) patients from \(F\) in stage
  2, and test both \(H_{0,S}\) and \(H_{0,F}\) at the final analysis.
\end{enumerate}

\begin{figure}
\centering
\includegraphics[width=\textwidth,height=8cm]{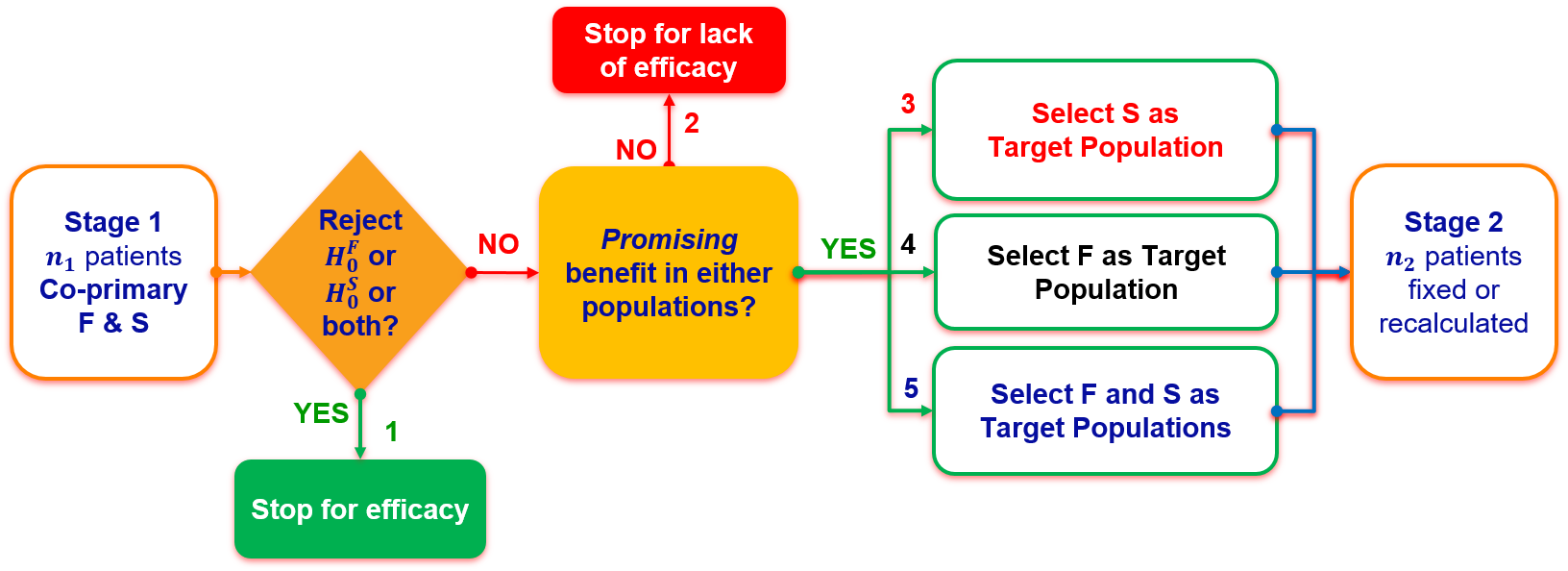}
\caption{Flow chart of adaptive enrichment design as defined above}
\end{figure}

Decision 1 provides an early opportunity to declare efficacy in case an
overwhelming benefit is observed in \(F\) or \(S\) (or both). Decision 2
avoids exposing additional patients to a potentially inefficacious
experimental treatment. The other three decisions are applicable if a
promising signal is seen but it is not pronounced enough for an early
read-out.

Decision 3 is chosen when the biomarker is strongly predictive of
treatment benefit, hence maximizing the power to reject \(H_0^S\) as
well as avoiding to expose patients in \(C\) to an ineffective treatment
with some potential safety reactions. Decisions 4 and 5 are appropriate
if a promising signal is observed in both \(S\) and \(C\) (or \(F\)).
Dropping \(S\) as a target population in Decision 4 is sensible when
e.g.~the predictive value of the relevant biomarker is weak, hence
dropping \(H_0^S\) and maximizing the probability to reject \(H_0^F\).

\hypertarget{type-i-error-control}{%
\subsection{Type I error control}\label{type-i-error-control}}

In an AED, type I error can be controlled by combining closed testing
\citep{Marcus76} with adaptive \(p\)-value combination. Denote the
unadjusted stage-wise \textcolor{blue}{one-sided} \(p\)-values
\textcolor{blue}{(testing for superiority of the intervention arm)}
corresponding to the null hypotheses \(H_0^q\) (\(q\in\{F,S\}\)) based
on data from stage \(i\) (\(i\in\{1,2\}\)) by \(p_i^q\).

\textcolor{blue}{Multiplicity in target populations is circumvented by using a closed testing procedure. For two target populations, the closed testing principle implies that significance in $S$ or $F$ can only be declared if the test of the intersection null hypothesis $H_0^S \cap H_0^F$ can also be rejected.  Several choices for the test of the intersection hypothesis are possible, see Section 11.2 of Wassmer and Brannath \citep{Wassmer2016a}. We use the Simes test \citep{Sanat_1997} because it protects type I error without requiring strong assumptions and is more powerful than the Bonferroni test. }

The design of the second stage of a two-stage AED, especially selection
of the target population(s) and associated test hypothesis(es), is
driven by trial data from the first stage which prohibits ``naive''
analyses via pooling data across stages. To remedy this, stage-wise
\(p\)-values can be combined by using inverse normal combination tests
\citep{lehmacher1999}. Define \(Z\)-values corresponding to stage 1
\(p\)-values by \(Z_1=\Phi^{-1}\left(1-p_1\right)\) for
\(p_1\in\{p_1^F,p_1^S,p_1^{(F,S)}\}\) and corresponding values \(Z_2\)
for stage 2 accordingly. A combined \(Z\)-value accross both stages is
then defined as \(\tilde{Z}_2=w_1 Z_1+w_2 Z_2\) with non-negative
weights \(w_1\) and \(w_2\) satisfying \(w_1^2+w_2^2=1\). The
\(p\)-value corresponding to \(\tilde{Z}_2\), that is,
\textcolor{blue}{$\tilde{p}_2=1-\Phi(\tilde{Z}_2)$}, is referred to as
the combined \(p\)-value.

The weights have to be pre-specified at the design stage. In our case
study, we use the common definition of the weights according to the
pre-planned stage-wise sample sizes \(n_1\) and \(n_2\):
\(w_1^2=n_1/(n_1+n_2)\) and \(w_2^2=n_2/(n_1+n_2)\). These weights are
optimal if the actual stage-wise sample sizes are proportional to the
planned stage-wise sample sizes in each population. Note that
discrepancies from this proportionality are expected in case the trial
is enriched at stage 2 and only subjects from \(S\) are recruited.
However, it has been shown that the associated power loss from this is
rather limited in all but extreme cases which is unlikely to our setting
\citep{lehmacher1999}.

Regardless of the adaptations after stage 1, \(Z_1\) and \(\tilde{Z}_2\)
follow the same bivariate distribution as a standard group sequential
test with two interim analyses at information fractions \(t_1=w_1^2\)
and \(t_2=1\). Thus, standard statistical software for group sequential
designs can be used for the determination of local significance levels
\(\alpha_1\) and \(\alpha_2\) after each stage which allow for early
stopping for efficacy and protect the ovarall significance levels across
both \(Z\)-tests.

\textcolor{blue}{Following the aforementioned close-test principle and combined p-value combination approach,}
overall test decisions which control the family-wise type I error in the
strong sense are then defined as follows:

\begin{itemize}
\tightlist
\item
  \(H_0^F\) is rejected after stage 1 if \(p_1^{(F,S)}\leq\alpha_1\) and
  \(p_1^{F}\leq\alpha_1\).
\item
  \(H_0^S\) is rejected after stage 1 if \(p_1^{(F,S)}\leq\alpha_1\) and
  \(p_1^{S}\leq\alpha_1\).
\item
  \(H_0^F\) is rejected after stage 2 if \(F\) is a target population in
  stage 2, with \(\tilde{p}_2^{(F,S)}\leq\alpha_2\) and
  \(\tilde{p}_2^{F}\leq\alpha_2\).
\item
  \(H_0^S\) is rejected after stage 2 if if \(S\) is a target population
  in stage 2, with \(\tilde{p}_2^{(F,S)}\leq\alpha_2\) and
  \(\tilde{p}_2^{S}\leq\alpha_2\).
\end{itemize}

Stopping for futility is discussed in Section 2.3.

\hypertarget{decision-criteria-determination-of-sample-size-and-other-design-parameters}{%
\subsection{Decision criteria, determination of sample size and other
design
parameters}\label{decision-criteria-determination-of-sample-size-and-other-design-parameters}}

If the AED cannot be stopped for compelling efficacy after stage 1, a
decision must be made whether to stop for futility or to continue to
stage 2 with one or both populations. Decision criteria may be based on
the observed treatment effect in \(S\) and \(C\) (and/or \(F\)) after
stage 1 (e.g. \citep{jenkins2010} and our case study), conditional or
predictive power (e.g. \citep{bhatt2016}), or Bayesian decision theory
(e.g. \citep{Goette2015,Kieser2015,Thomas2017}). While type I error
control is guaranteed regardless of how these choices are made, the
decision criteria affect the probability of correct decisions after
stage 1 and study power.

In general, the design parameters for an AED include (1) the sample
sizes \(n_1\) and \(n_2\) of stage 1 and 2, rescectively, where \(n_2\)
could be re-calculated at the interim analysis, (2) the
\(\alpha\)-spending approach for early stopping for efficacy, (3) the
exact decision criteria (thresholds) used for population selection
criteria as well as for early stopping for futility.

In practice, these design parameters are usually determined by running
simulations across a range of plausible scenarios and evaluating design
characteristics such as: the overall power of the study, that is, the
probability of a statistically significant result for either of the
target populations or both at either the interim or final analysis, the
conditional power, that is, the probability of a significant result
conditional on continuing to stage 2, and the probability of making the
``correct'' decision(s) at the interim analysis. Overall power measures
the success probability of the entire AED whereas conditional power
assesses the probability of success of the additional investment into
stage 2. Usually, a trade-off between overall and conditional power
needs to be made because maximizing the latter leads to aggressive
thresholds for futility stopping and population selection which may
reduce overall power.

\textcolor{blue}{Of note, often ``biomarker status'' could be used as stratification factor for randomization and analysis. We adopt a common practice, which is to neglect this in sample size calculation for the case study in this paper.}

\hypertarget{minimal-detectable-difference}{%
\subsection{Minimal detectable
difference}\label{minimal-detectable-difference}}

In Section 2.2, it was outlined how to determine local significance
levels \(\alpha_1\) and \(\alpha_2\), that is, boundary values based on
which test decisions can be taken as described at the end of Section
2.2.

To support clinical interpretation of these local significance levels,
it is useful to express them on the treatment effect scale. We denote
this as the minimal detectable difference (MDD), that is, the smallest
observed absolute risk difference between the two groups which would
lead to rejection of the corresponding null hypothesis after stage 1 or
2, respectively.

MDDs (also called boundary values on the treatment effect scale) are
routinely provided for single stage or group-sequential trials by
standard software such as rpact \citep{rpact} (version 2.0.5) and an
extension to AEDs is described below.

We first consider a single stage trial and a statistical test of the
null hypothesis \(H_0:\Delta=\pi_1-\pi_2\le0\) versus the alternative
\(H_A:\Delta=\pi_1-\pi_2>0\) at the one-sided significance level
\(\alpha^*\). The observed proportion of \(\hat{\pi}_2\) in the control
arm serves as a nuisance parameter in this setting and, typically, it is
assumed to correspond to the hypothesized control proportion from the
sample size calculation. Given \(\hat{\pi}_2\), the MDD \(\delta^*\)
corresponds to the observed difference between the two arms which would
lead to a \(p\)-value of exactly \(\alpha^*\) using e.g.~a signed
(one-sided) chi-squared test for hypothesis testing. Numerically, the
MDD can be calculated with any one dimensional root (zero) finding
algorithm.
\textcolor{blue}{If there is substantial uncertainty regarding the true response probability in the control arm, the MDD should be calculated for a range of plausible values.}

In an AED, MDDs which lead to the rejection of the respective population
null hypothesis \emph{after stage 1} can be calculated in the same way
as for a single stage trial. The only additional complication is that
the corresponding null hypothesis for each population can only be
rejected if the intersection hypothesis is also rejected. If Simes test
is used to test the intersection hypothesis, the intersection hypothesis
and the null hypothesis for \(S\) can both be rejected after stage 1 if
either \(2p_1^S\leq\alpha_1\) (that is, \(S\) alone is responsible for
the rejection of the intersection null hypothesis) or if both
\(p_1^S\leq \alpha_1\) and \(p_2^F\leq\alpha_1\) (that is, \(F\)
contributes to the rejection of the intersection null hypothesis).
Consequently, two MDDs can be calculated for \(S\): First, a
conservative MDD for \(S\) which assumes that \(S\) alone is responsible
for rejection of the intersection null hypothesis and corresponds to the
MDD for a single stage trial based on the ``adjusted'' \(p\)-value
\(2p_1^S\). Second, a more liberal MDD which is only applicable if \(F\)
contributes to the rejection of the intersection null hypothesis (and
hence is also significant with a smaller \(p\)-value than \(S\)), and
corresponds to the MDD for a single stage trial based on the raw
\(p\)-value \(p_1^S\). Both, the conservative and liberal MDD will be
useful in design discussions with the drug development team. In an
analogous way, two MDDs can be calculated for \(F\).

MDD calculations \emph{after stage 2} are based on the combination test
and require additional consistency assumptions to achieve a unique
solution. First, we assume that the estimated proportions in each arms
are identical for stage 1 and stage 2 data, respectively.

If \emph{only \(S\) continues to stage 2}, the intersection test is only
relevant for the stage 1 data and it is plausible to additionally assume
consistency in the driver for the treatment effect, that is, that the
stage 1 \(p\)-value for Simes test is driven by \(S\) alone and given by
\(2p_1^S\). Thus, the MDD for \(S\) after stage 2 corresponds to the
observed difference between the two arms which would lead to a
\(p\)-value of exactly \(\alpha_2\) using the combination test based on
the adjusted stage 1 \(p\)-value \(2p_1^S\) and the raw stage 2
\(p\)-value \(p_2^S\). If \emph{only \(F\) continues to stage 2}, the
MDD can be calculated in an analogous way.

If \emph{both \(S\) and \(F\) continue to stage 2}, we propose to
calculate a conservative MDD for \(S\) using an adjusted stage 1
\(p\)-value \(2p_1^S\) and an adjusted stage 2 \(p\)-value \(2p_2^S\)
for the combination test (assuming that the intersection test is driven
by \(S\) alone in both stages) and a more liberal MDD for \(S\) based on
the raw stage 1 \(p\)-value \(p_1^S\) and the raw stage 2 \(p\)-value
\(p_2^S\) (which is only valid if \(F\) contributes to the rejection of
the intersection test in both stages). In the same way, a conservative
and a liberal MDD can be calculated for \(F\).

As described above, MDDs can only be derived under additional
assumptions. While these assumptions will often be approximately true,
it is important to recognize that formal test decisions should be based
on the local significance levels \(\alpha_1\) and \(\alpha_2\) and not
on the MDD. However, the MDD is on a clinically relevant scale, and as
such extremely helpful for a discussion of the trial design with the
cross-functional clinical development teams. For example, if the MDD is
very small, this indicates that it is possible that the observed
treatment effect is not clinically relevant but that the trial is
nevertheless statistically significant. This would make it very
difficult to market the drug and hence, such a finding may lead to a
re-consideration of the trial sample size.

\hypertarget{application-the-impassion031-trial}{%
\section{Application: the IMpassion031
trial}\label{application-the-impassion031-trial}}

\hypertarget{the-original-impassion031-trial-design}{%
\subsection{The original IMpassion031 trial
design}\label{the-original-impassion031-trial-design}}

IMpassion031 is a global Phase III, double-blind, 1:1 randomized,
multicenter, placebo-controlled study which is conducted to evaluate the
efficacy and safety of neoadjuvant treatment with nab-paclitaxel +
doxorubicin + cyclophosphamide and either atezolizumab or placebo in
invasive stage II/III early triple-negative breast cancer (TNBC). The
CIT atezolizumab is an anti-programmed death-ligand 1 (PD-L1) monoclonal
antibody that blocks the binding of PD-L1 to PD-1 and B7.1 receptors,
thereby restoring tumor-specific immunity. The primary efficacy endpoint
of IMpassion031 is pathological complete response (pCR), a binary
clinical endpoint evaluated at surgery which takes place approximately 6
months after randomization for patients in both arms. pCR in this study
is defined as absence of residual invasive cancer in the complete
resected breast specimen and all sampled regional lymph nodes following
completion of neoadjuvant therapy \citep{FDA_pCR}. The original
IMpassion031 trial had a fixed non-adaptive design with
\textcolor{blue}{a one-sided significance level of 2.5\% and} a target
sample size of 204 subjects from the overall population \(F\),
randomized 1:1 to receive either atezolizumab in combination with
chemotherapy or chemotherapy alone. This yields a power of 79\% to
detect an increase of 20\% pCR rate in the combo arm from a true pCR
rate of 48\% in the mono arm, accounting for 5\% drop-out rate in both
treatment arms, whereby drop-outs are considered non-responders.
\textcolor{blue}{Chi-square test for two proportions was considered.}

The trial was initiated on 24th July 2017 and recruitment of the
original 204 subjects independent of PD-L1 status was completed on 12th
June 2018, with 205 patients actually enrolled. During patient follow-up
and prior to study unblinding, data external to the trial emerged
suggesting that PD-L1 could be a predictive biomarker for the treatment
effect of atezolizumab in metastatic or locally advanced TNBC
(IMpassion130 \citep{IMpassion130_Press}). Although results of
IMpassion130 were compelling with respect to the predictive nature of
PD-L1 biomarker, the extent to which this finding would apply to the
early TNBC setting was uncertain. Therefore, the original fixed design
was changed to an AED to address this potential predictive biomarker
population hypothesis. PD-L1 status was dichotomized according to a
pre-specified cut-point which has been previously used and validated in
the pivotal trial in advanced TNBC \citep{IMpassion130}.

\hypertarget{the-impassion031-aed}{%
\subsection{The IMpassion031 AED}\label{the-impassion031-aed}}

By the time of the protocol amendment, that is, when transforming the
trial to an AED, the target sample size for the original design had
already been fully enrolled and follow-up was ongoing. Thus, this phase
of the trial was assigned as stage 1 of the new AED with \(n_1=205\)
subjects.

The new AED design has the following features. The overall one-sided
type-I error level is \(\alpha=2.5\)\% and both used a 1:1 randomization
ratio. After the interim analysis at the end of stage 1, to safeguard
the scientific integrity of the study, an independent data monitoring
committee will look at unblinded data to make recommendations to the
trial sponsor regarding early stopping for efficacy or futility or
continuing into stage 2 with the target population(s) selected,
following the general framework laid down in Section 2 and displayed in
Figure 1. The sample size of \(n_2=120\) subjects for stage 2 is
determined based on its favorable design characteristics such as power
(see below) without adding too much operational complication such as
extension of study duration. Importantly, the protocol amendment was
completed and submitted to regulatory authorities while the treatment
assignment remained double-blinded, hence protecting the integrity of
the trial.

Type I error control is implemented by combining closed testing via
Simes test with \(p\)-value combination using inverse normal combination
tests as described in Section 2. In order to allow for an acceptably
high probability that the trial stops after stage 1, that is, the sample
size of the original trial, 50\% of the overall type I error is spent at
stage 1 which implies local significance levels of \(\alpha_1=0.0125\)
and \(\alpha_2=0.0184\).

Adaptive sample size re-calculation after stage 1 is not considered
since the design with fixed stage 2 sample has sufficient power and
allowed to control the overall trial timelines. The selection of target
population(s) for stage 2 and futility decisions are based on the
observed treatment effect, that is, the difference of pCR rates between
the two arms, after stage 1 in the PD-L1 positive population \(S\) and
its counterpart, the PD-L1 negative population \(C\). Based on clinical
trial simulations (reported below) and discussion with clinicians,
futility decision thresholds \(d_S=12\)\% and \(d_C=10\)\% are chosen.
Thus, if the observed treatement effect in the PD-L1 positive population
is \(\ge12\)\% after stage 1, it is included as a target population in
stage 2. Similarly, if the observed treatment effect in the PD-L1
negative population is \(\ge10\)\% after stage 1, the full population
\(F\) is included as a target population in stage 2. If both treatment
effects exceed the threshold, both populations are included and tested
as co-primary populations at the final analysis. If neither treatment
effect is above the threshold, the trial stops for futility. Decisions
concerning \(F\) are predicated on the observed treatment in the PD-L1
negative population instead of the full population to avoid cases where
the benefit in \(F\) is strongly driven by \(S\), hence minimizing the
risk of exposing patients in \(C\) to a potentially futile treatment
independent of its activity in \(S\).

Simulations are used to investigate the properties of the chosen AED and
the decision thresholds \(d_S\) and \(d_C\). Three scenarios are
reported here which all assume a prevalence of 47\% of the PD-L1
positive subgroups
\textcolor{blue}{(based on internal and external information such as Kwa et. al \citep{Kwa2018})},
a pCR rate of 48\% in the control arm, and a treatment effect of 20\% in
the PD-L1 positive subgroup. The treatment effect in PD-L1 negative
subjects is varied between 4\%, 12\%, and 20\% (Table 1). The original
trial design assumed a homogenous treatment effect of 20\% in the full
population but scenarios with a reduced treatment effect in PD-L1
negative patients are also plausible in view of the external information
that triggered the amendment to IMpassion031. Of note, all simulations
assumed a drop-out rate of 5\% in both treatment arms and stages
independent of the pCR outcome and that drop-outs were considered
non-responders.
\textcolor{blue}{Therefore, as an example, an underlying control arm pCR of 0.48$\times$0.95=0.456 is considered in our simulation''}.

Results of the simulations are displayed in Tables 2-4. In addition to
the chosen decision thresholds of \(d_S=12\)\% and \(d_C=10\)\%, results
for an alternative set of more aggressive decision thresholds
\(d_S'=15\)\% and \(d_C\)'=12\% are also displayed for comparison
purposes. Reported results for each scenario represent averages over
100,000 simulated trials.

Under the original trial assumptions (scenario 1), the overall
probability to stop for efficacy (that is, reject either \(H_{0,S}\) or
\(H_{0,F}\) or both) after stage 1 is 64\% compared to a power of 79\%
for the original trial design (Table 2). This quantifies the price for
the additional flexibility of two co-primary populations, the choice of
the target population for stage 2, and a second opportunity to declare
efficacy after stage 2. The futility decision thresholds \(d_S\) and
\(d_C\) do not impact the estimated probabilities of stopping for
efficacy at stage 1. However, the more aggressive (higher) decision
thresholds almost double the likelihood for futility stopping in all
scenarios, whereas they reduce the likelihood of continuing to stage 2
with only patients from \(S\). Similarly, using the more aggressive
thresholds reduce the chance of continuing to stage 2 with both \(S\)
and \(F\) whereas the chance of continuing only in \(F\) is slightly
increased in all scenarios. This latter finding is mainly due to the
fact that many of the simulated trials which would continue to stage 2
with both \(F\) and \(S\) for the less aggressive thresholds continue
with only \(F\) for the more aggressive thresholds due to the larger
increase from \(d_S\) to \(d_S'\) compared to the increase from \(d_C\)
to \(d_C'\). Hence, \(d_S=12\)\% and \(d_C=10\)\% are considered the
better choice due to the lower chance of stopping for futility as well
as higher chance to continue in either \(S\) or both \(F\) and \(S\),
which provides more confidence to clinicians based on the mode of action
of our active treatment.

The overall power for the revised IMpassion031 design is 88\% (versus
79\% for the original design) for scenario 1, 76\% (versus 57\%) for
scenario 2, and 67\% (versus 35\%) for scenario 3 (Table 3). While the
power is still below 80\% for scenarios with a reduced treatment benefit
in PD-L1 negative patients, it is substantially increased compared to
the original design. An even larger increase in power would have
required a larger stage 2 sample size, which would have put challenges
on recruitment feasibility and timelines, in particular if one would
only continue with \(S\) in stage 2.

More aggressive thresholds for assessing futility and population
selection would lead to a general drop in overall power (Table 3) but
increase the conditional power, that is, the conditional rejection
probabilities if stage 2 is activated (Tables 4). However, one needs to
take into account the greater risk of not activating stage 2 with the
more aggressive thresholds.

Finally, MDDs for IMpassion 031 are displayed in Table 5. An observed
difference in response rates of at least 17\% in \(F\) or at least 25\%
in \(S\) (or a difference of both at least 16\% in \(F\) and at least
22\% in \(S\) jointly) is required for an efficacy stop after stage 1.
After stage 2, the smallest MDD is 12\% for \(F\) which applies if \(F\)
and \(S\) are both tested at stage 2 and the treatment effect in \(S\)
alone is also at least 17\%. These MDDs were all considered to
correspond to clinically relevant effect sizes.

\textcolor{blue}{Of note, the futility boundaries at the interim were $d_S=12$\% and $d_C=10$\% (consistent with a treatment effect of $\approx 11$\% in $F$). These values are smaller than the calculated  MDDs after stage 2. If this were not the case, the same observed treatment effect might lead to a futility stop at the interim analysis but also to a rejection of the null hypothesis at the final analysis. This would be incoherent and should lead to a re-assessment of the planned futility boundaries.}

\hypertarget{discussion}{%
\section{Discussion}\label{discussion}}

The world of cancer immunotherapies (CITs) is highly dynamic and data
external to an ongoing trial is evolving quickly. In this paper, we
showed that it is possible to incorporate emerging data regarding the
most appropriate target population into an ongoing trial by amending it
to an adaptive enrichment trial (AED). The implemented two-stage AED
allows for both efficacy and futility stopping after stage 1, as well as
for population selection for stage 2. Importantly, the protocol was
amended and submitted to regulatory agencies while stage 1 was still
ongoing and prior to any unblinding of the double-blind trial, hence
protecting the integrity of the design.

This paper also introduced the concept of the minimal detectable
difference (MDD) to AED, which is relevant for the discussion of the
design with stakeholders beyond biostatistics.

The proposed AED in CIT addresses several important considerations
highlighted in the recent adaptive trials guidance by the FDA for
enrichment designs \citep{FDA_AD}: First, the design combines
established statistical methods to protect the familywise type I error
in the strong sense. Second, the data external to the trial provided a
strong rationale that the benefit-risk profile may be more favorable in
the PD-L1 positive subgroup. Third, the PD-L1 assay and the threshold
used to define PD-L1 positivity had been previously validated
\citep{IMpassion130}. Fourth, stage 1 of the trials was large enough to
characterize the treatment effect in the complement population, that is,
the PD-L1 negative subgroup, even in the situation when only PD-L1
positive subjects would be included in stage 2.

However, our proposal also has several limitations. The amendment was
promptly implemented following the release of the external data, but
this occurred only after recruitment to stage 1 of the trial had already
been completed. This prevented optimization of some design parameters of
the AED such as the stage 1 sample size. Moreover, our derivation of the
MDD relied on the simple dependence of Simes' test on the p-values from
each population. If more complex intersection tests are employed,
e.g.~the test proposed by Spiessen and Debois \citep{spiessens2010},
this would further complicate the quantification of the dependence of
test decisions for each population on the intersection test.
\textcolor{blue}{In addition, we focused on hypothesis testing but did not cover estimation and inference in AEDs which is an important area of current research. We refer to Chapter 8 of Wassmer and Brannath \citep{Wassmer2016a} for a general discussion for adaptive trials and to Kunzman et al \citep{Kunzmann2017} for a discussion of estimation in the context of AEDs.}
Finally, the current setting of a binary endpoint is methodologically
easier than the time-to-event setting, which is frequent in oncology,
where additional complications arise \citep{jenkins2010,Dominic2016}.

In summary, this case study demonstrated that AED are an efficient way
to circumvent emerging uncertainties about the target population for
cancer immunotherapy. AEDs are still relatively rarely used in clinical
development and we hope that this paper promotes their use and increases
the confidence that such designs are feasible in practice.

\hypertarget{software}{%
\section{Software}\label{software}}

The code and the results of all computations described in this paper are
available as a GitHub repository:
\url{https://github.com/nguyenducanhvn101087/Enrichment_Adaptive_Design}

\hypertarget{data-availability-statement}{%
\section{Data Availability
Statement}\label{data-availability-statement}}

Data sharing is not applicable to this article as no new data were
created or analyzed in this study.

\bibliographystyle{unsrtnat}

\bibliography{aed}

\newpage

\hypertarget{tables}{%
\section{Tables}\label{tables}}

Table 1: Scenarios investigated in the clinical trial simulations

\begin{table}[H]
\centering
\begin{threeparttable}
\begin{tabular}{lllll}
\toprule
Scenario & $\pi_1^S-\pi_2$ & $\pi_1^C-\pi_2$ & $\pi_1^F-\pi_2$ & $\pi_2$\\
\midrule
1 & 0.20 & 0.20 & 0.20 & 0.48\\
2 & 0.20 & 0.12 & 0.16 & 0.48\\
3 & 0.20 & 0.04 & 0.12 & 0.48\\
\bottomrule
\end{tabular}
\begin{tablenotes}
\small
\item [] $\pi_1^q$ is the pCR rate in intervention arm for population $q$, $q\in(F,S,C)$. $\pi_2$ is the pCR rate in control arm assumed to be the same for $S$ and $C$. Numbers in the table refer to true pCR rates.
\end{tablenotes}
\end{threeparttable}
\end{table}

~ ~

Table 2: Relative Frequencies of Decisions at Stage 1 for IMpassion031
for the different scenarios and different futility decision thresholds

\begin{table}[H]
\centering
\begin{tabular}{lllllll}
\toprule
\multicolumn{1}{c}{ } & \multicolumn{3}{c}{$d_S=0.12, d_C=0.1$} & \multicolumn{3}{c}{$d_S=0.15, d_C=0.12$} \\
\cmidrule(l{3pt}r{3pt}){2-4} \cmidrule(l{3pt}r{3pt}){5-7}
Decision & Scen. 1 & Scen. 2 & Scen. 3 & Scen. 1 & Scen. 2 & Scen. 3\\
\midrule
1a - Stop after stage 1: efficacy in F only & 0.27 & 0.12 & 0.04 & 0.27 & 0.12 & 0.04\\
1b - Stop after stage 1: efficacy in S only & 0.01 & 0.04 & 0.10 & 0.01 & 0.04 & 0.10\\
1c - Stop after stage 1: efficacy in F and S & 0.36 & 0.29 & 0.19 & 0.36 & 0.29 & 0.19\\
2 - Stop after stage 1: futility & 0.04 & 0.10 & 0.17 & 0.08 & 0.19 & 0.29\\
3- Continue to stage 2: S only & 0.08 & 0.22 & 0.38 & 0.07 & 0.18 & 0.29\\
4- Continue to stage 2: F only & 0.14 & 0.11 & 0.06 & 0.17 & 0.14 & 0.06\\
5- Continue to stage 2: F and S & 0.10 & 0.11 & 0.07 & 0.03 & 0.04 & 0.02\\
\bottomrule
\end{tabular}
\end{table}

~ ~

Table 3: Overall statistical power for IMpassion031 for the different
scenarios and different futility decision thresholds

\begin{table}[H]
\centering
\begin{tabular}{lllllll}
\toprule
\multicolumn{1}{c}{ } & \multicolumn{3}{c}{$d_S=0.12, d_C=0.1$} & \multicolumn{3}{c}{$d_S=0.15, d_C=0.12$} \\
\cmidrule(l{3pt}r{3pt}){2-4} \cmidrule(l{3pt}r{3pt}){5-7}
  & Scen. 1 & Scen. 2 & Scen. 3 & Scen. 1 & Scen. 2 & Scen. 3\\
\midrule
Power F & 0.80 & 0.54 & 0.28 & 0.79 & 0.52 & 0.27\\
Power S & 0.49 & 0.57 & 0.61 & 0.45 & 0.51 & 0.55\\
Power (F or S) & 0.88 & 0.76 & 0.67 & 0.86 & 0.71 & 0.61\\
Power (F and S) & 0.41 & 0.35 & 0.22 & 0.38 & 0.32 & 0.20\\
\bottomrule
\multicolumn{7}{l}{\textsuperscript{} Power (F or S) considers statistical significance achieved for either F}\\
\multicolumn{7}{l}{or S or both}\\
\end{tabular}
\end{table}

~ ~

Table 4: Conditional power (CP): Power conditional on activation of
stage 2 for IMpassion031 for the different scenarios and different
futility decision thresholds

\begin{table}[H]
\centering
\begin{tabular}{lllllll}
\toprule
\multicolumn{1}{c}{ } & \multicolumn{3}{c}{$d_S=0.12, d_C=0.1$} & \multicolumn{3}{c}{$d_S=0.15, d_C=0.12$} \\
\cmidrule(l{3pt}r{3pt}){2-4} \cmidrule(l{3pt}r{3pt}){5-7}
  & Scen. 1 & Scen. 2 & Scen. 3 & Scen. 1 & Scen. 2 & Scen. 3\\
\midrule
CP F if only F tested in stage 2 & 0.67 & 0.46 & 0.27 & 0.74 & 0.57 & 0.40\\
CP S if only S tested in stage 2 & 0.77 & 0.76 & 0.74 & 0.84 & 0.83 & 0.82\\
CP (F or S) 
 if F and S tested in stage 2 & 0.82 & 0.72 & 0.61 & 0.86 & 0.79 & 0.68\\
\bottomrule
\end{tabular}
\end{table}

~ ~

Table 5: Minimal detectable differences (MDDs) for the IMpassion031
trial, that is, approximate minimum absolute risk differences that would
lead to a significant result in that population after the respective
stage.

\begin{table}[H]
\centering
\begin{tabular}{lll}
\toprule
Stage & Subgroup S & Full population F\\
\midrule
Stage 1 & 0.25 [0.22] & 0.17 [0.16]\\
Stage 2 &  & \\
- only S tested in stage 2 & 0.16 & \\
- only F tested in stage 2 &  & 0.13\\
- F and S included in stage 2 & 0.20 [0.17] & 0.14 [0.12]\\
\bottomrule
\multicolumn{3}{l}{\textsuperscript{} All MDDs assume a pCR probability of 0.48$\times$0.95=0.456 in}\\
\multicolumn{3}{l}{the control arm accounting for 5\% drop-outs. Numbers in}\\
\multicolumn{3}{l}{brackets are valid if the respective other population drives the}\\
\multicolumn{3}{l}{intersection test, that is, both populations statistically}\\
\multicolumn{3}{l}{significant but the respective other population has a lower}\\
\multicolumn{3}{l}{$p$-value.}\\
\end{tabular}
\end{table}

\end{document}